\documentclass[%
 reprint,
groupedaddress,
 amsmath,amssymb,
 aps,
prl,
floatfix,
]{revtex4-2}
\usepackage{graphicx}
\usepackage{dcolumn}
\usepackage{bm}
\usepackage{color}

\begin{document}

\preprint{APS/123-QED}

\title{Broadband amplitude squeezing in electrically driven quantum dot lasers}
\author{Shiyuan Zhao}
 \email{shiyuan.zhao@telecom-paris.fr}
\author{Shihao Ding}
 \email{shihao.ding@telecom-paris.fr}
\author{Heming Huang}
\author{Isabelle Zaquine}
\author{Nicolas Fabre}
\author{Fr\'ed\'eric Grillot}
 \affiliation{%
 LTCI, T{\'e}l{\'e}com Paris, Institut Polytechnique de Paris, 19 Place Marguerite Perey, 91120, Palaiseau, France
}
\author{Nadia Belabas}
\affiliation{
    Universit{\'e} Paris-Saclay, CNRS, Centre de Nanosciences et de Nanotechnologies, 91120, Palaiseau, France
}

\date{\today}

\begin{abstract}
The generation of broadband squeezed states of light lies at the heart of high-speed continuous-variable quantum information. Traditionally, optical nonlinear interactions have been employed to produce quadrature-squeezed states. However, the harnessing of electrically pumped semiconductor lasers offers distinctive paradigms to achieve enhanced squeezing performance. We present evidence that quantum dot lasers enable the realization of broadband amplitude-squeezed states at room temperature across a wide frequency range, spanning from 3 GHz to 12 GHz. Our findings are corroborated by a comprehensive stochastic simulation in agreement with the experimental data. 
\end{abstract}
                            
\maketitle

The evolution of photonics-based quantum information technologies is currently on the brink of initiating a revolutionary transformation in data processing and communication protocols \cite{yin2017satellite,liao2017satellite}. A cornerstone within this realm will be the quantum emitter. In recent years, there has been a substantial upsurge in both theoretical and experimental investigations centred around semiconductor quantum dot (QD) nanostructures \cite{michler2017quantum}. A particular emphasis has been placed on self-assembled QDs embedded into microcavities, which facilitate the generation of single photons with high purity and indistinguishability \cite{michler2000quantum, somaschi2016near,arakawa2020progress}. As a result, such sources assume a pivotal role in quantum computing \cite{wang2019boson,coste2023high} as well as the discrete variables (DV) quantum key distribution (QKD) \cite{vajner2022quantum}. In stark contrast to the DV QKD, which requires single-photon sources and detectors, continuous variable (CV) QKD leverages lasers and balanced detection to continuously retrieve the light's quadrature components during key distillation. This approach benefits from readily available equipment and seamless integration into existing optical telecommunications networks \cite{zhang2020long}. One of the CV QKD protocols, GG02 \cite{grosshans2002continuous}, is widely acclaimed for its security due to the no-cloning theorem of coherent states \cite{glauber1963coherent}. Nevertheless, a recent study has delved into the use of squeezed states to achieve even higher levels of security and robustness \cite{jacobsen2018complete}. This innovative approach strives to completely eliminate information leakage to potential eavesdroppers in a pure-loss channel and to minimize it in a symmetric noisy channel. Within this cutting-edge protocol, information can be exclusively encoded through a Gaussian modulation of amplitude-squeezed states, which are commonly referred to as photon-number squeezed states. These states demonstrate reduced fluctuations in photon number $\left<\Delta\hat{n}^2\right><\left<\hat{n}\right>$ with respect to coherent states, albeit encountering enhanced phase fluctuations due to the minimum-uncertainty principle. 
\begin{figure*}[htbp]
\centerline{\includegraphics[width=1\textwidth]{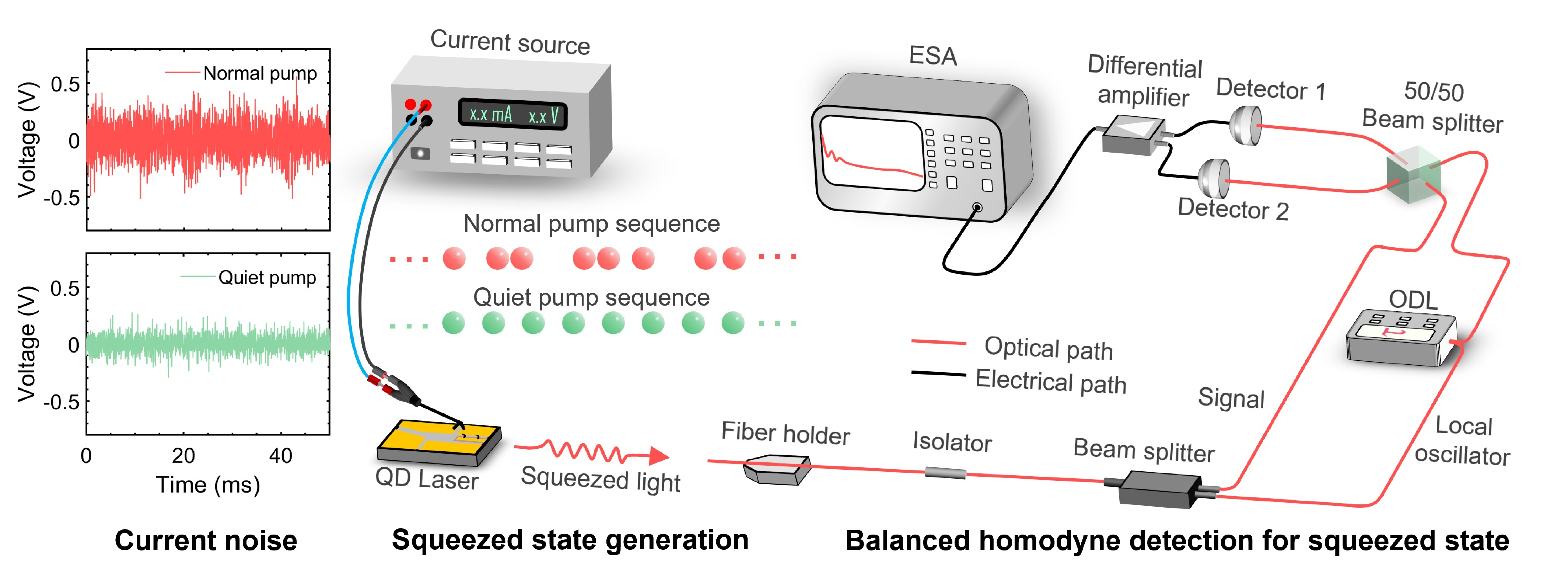}}
\caption{Diagram of the key components of the apparatus used for generating amplitude-squeezed states through QD lasers. Left panel: Comparison of noise characteristics between two different current sources, namely Keithley 2400 (normal pump, red) and ILX Lightwave LDX-3620 (quiet pump, green). The current noises are assessed by detecting voltage fluctuations across a 10 $\Omega$ resistor. Further details can be found in the Supplementary Material. Right panel: Experimental setup. The quiet pump injects electrons at a constant rate. ODL is the optical delay line and ESA is the electrical spectrum analyzer. \label{Fig:1}}
\end{figure*}

Over the past years, squeezed states of light have been frequently generated using $\chi^{(2)}$ or $\chi^{(3)}$ nonlinear interactions via parametric down-conversion and four-wave mixing \cite{slusher1985observation,*wu1986generation}. A variety of nonlinear materials have been applied in these processes, including LiNbO$_3$ (PPLN) \cite{kashiwazaki2023over}, KTiOPO$_4$ (PPKTP) \cite{vahlbruch2016detection}, silicon \cite{safavi2013squeezed}, atomic vapour \cite{mccormick2007strong}, disk resonator \cite{furst2011quantum}, and Si$_3$N$_4$ \cite{zhao2020near}. Recent advancements have also facilitated the transition from traditional benchtop instruments to a more compact single-chip design \cite{dutt2015chip,mondain2019chip,moody2020chip,yang2021squeezed,nehra2022few}. As opposed to that, Y. Yamamoto \textit{et al.}~\cite{yamamoto1986amplitude,*machida1987observation} initially proposed an alternative to produce amplitude-squeezed states directly with off-the-shelf semiconductor lasers using a "quiet" pump, i.e. a constant-current source. A striking peculiarity of semiconductor lasers is their ability to be pumped by injection current supplied via an electrical circuit. Unlike optical pumping, electrical pumping is not inherently a Poisson point process due to the Coulomb interaction and allows for reducing pump noise below the shot-noise level \cite{yamamoto1986amplitude}. Notably, this method can take full advantage of the mature fabrication processes in the semiconductor industry, thereby significantly boosting its feasibility. In other words, its efficacy hinges on the improved performance of recently developed semiconductor lasers, characterized by their compact footprint, ultralow intensity noise and narrow frequency linewidth. While subsequent experiments involving various types of laser diodes have gained widespread interest in this domain, including commercial quantum well (QW) lasers \cite{wang1993squeezed,kitching1995room}, vertical-cavity surface-emitting lasers \cite{kaiser2001amplitude}, and semiconductor microcavity lasers \cite{choi2006observation}, the observed bandwidth has, until now, remained relatively limited. The broadest achieved bandwidth has reached 1.1 GHz with a QW transverse junction stripe laser operating at a cryogenic temperature of 77 K \cite{machida1988ultrabroadband}. This limitation indeed poses strict constraints on the practical implementation of room-temperature conditions and hinders the realization of high-speed quantum communications. While a recent study did anticipate the theoretical potential of producing broadband amplitude-squeezed states in interband cascade lasers \cite{zhao2022stochastic}, it is worth noting that, prior to this Letter, no experimental demonstration of such phenomenon had been presented.

In this Letter, we unveil experimental evidence of the generation of broadband amplitude-squeezed states from a QD semiconductor laser chip. Departing from the conventional QW active region, QD lasers achieve complete spatial quantization, resulting in unparalleled three-dimensional carrier confinement \cite{grillot2021uncovering}. Our results confirm that a constant-current-driven QD laser at room temperature can generate sub-Poissonian light with an extended bandwidth of 9 GHz. This outcome closely aligns with Yamamoto's theoretical framework \cite{yamamoto1986amplitude}, which suggests that the intensity noise of the laser output field, rather than the internal field, may be diminished below the shot-noise limit at the high pump current. The shot noise limit originates from two sources, namely, pump fluctuations in the low-frequency region and reflected vacuum-field fluctuations in the high-frequency region \cite{yamamoto1986amplitude}. In order to eliminate pump current fluctuations, we effectively implemented the high-impedance configuration to achieve uniform regulation of the pumping electrons \cite{machida1987observation} and accurately characterized the noise power of the current sources. Moreover, we have corroborated our findings through numerical simulations that exhibit a good agreement with experimental data.

A diagram of the experimental setup is illustrated in Figure.~\ref{Fig:1}. A distributed-feedback (DFB) single-mode QD laser is utilized, emitting at an oscillation wavelength of 1.31 $\mu$m. The laser features a highly reflective (HR) coating (95\%) on the rear facet and an antireflection (AR) coating (3\%) on the front facet. To maintain a constant operating temperature of 20 {\textcelsius}$\pm$0.005{\textcelsius}, a thermoelectric temperature controller (ILX Lightwave LDT-5748) is thus applied. The QD laser's threshold current $I_{th}$ is measured to be approximately 9 mA. The primary factor contributing to the modest differential quantum efficiency of 20\% is the loss incurred through output fibre coupling. Nevertheless, this does not exert an impact on the internal quantum efficiency that remains close to unity and pertains to the conversion from injected electrons to emitted photons. The stable single-longitudinal-mode operation is realized over a wide pump current range and the side mode suppression rejection (SMSR) is more than 50 dB at the high pump currents to avoid the side-mode-competition interference \cite{marin1995squeezing}. 

\begin{figure}[htbp]
\centerline{\includegraphics[width=0.5\textwidth]{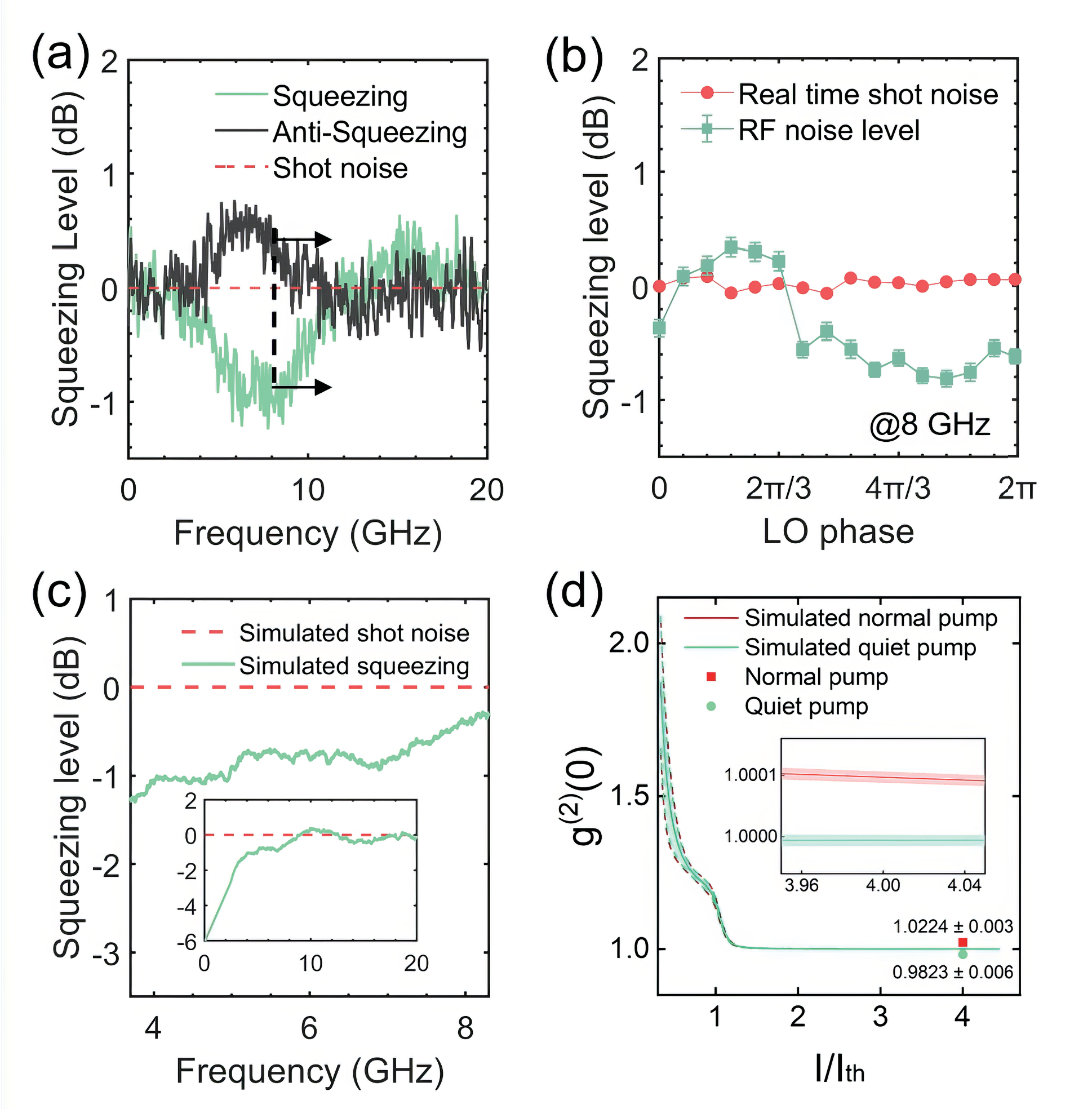}}
\caption{(a) The measured RF spectra at 40 mA are represented by the green line (squeezing) and black line (anti-squeezing). When the LO branch is blocked, the vacuum field entering the signal port of the detector produces the vacuum noise level shown by the red dashed line, which remains insensitive to changes in $\phi_{LO}$. The spectrum analyzer was set to a resolution bandwidth (RBW) of 200 kHz and a video bandwidth (VBW) of 500 Hz. All traces have been corrected for thermal background noise subtraction. (b) LO phase dependence of the quantum fluctuations in the amplitude-squeezed state produced by a constant-current-driven QD laser at 8 GHz. With the LO branch present but the LO phase varying, the calibrated SQL demonstrates great stability with a small deviation of $\pm$0.08 dB. The error bar for the green point was obtained by averaging the noise curve over a range of $\pm$50 MHz around 8 GHz. (c) The stimulated RF spectra are shown in green, while the stimulated SQL is represented in red dashed line. (d) The stimulated normalized second-order correlation function at zero delay $g^{(2)}(0)$ as a function of pump current. The green and grep areas show the standard deviation from the average obtained from 100 simulation runs. Two experimental data points are also presented. \label{Fig:2}}
\end{figure}

The squeezed light is measured with a standard balanced homodyne detection \cite{yuen1983noise}. For the measurement process, the DFB QD laser is coupled into a fibre holder and precisely driven by a low-noise current source (ILX Lightwave LDX-3620). A 30 dB optical isolator is imperative because even a very small amount of back-reflected light leads to excessive intensity noise. After the optical isolator, the output beam is split into two separate paths by a 90/10 beamsplitter. One path is dedicated to detecting the squeezed signal, while the other is used for the strong local oscillator (LO). The LO serves as a phase reference, and by scanning the signal with an optical delay line, it can effectively shift the relative phase, enabling the distinction between the field quadratures. The two paths are then recombined into a tunable beamsplitter to ensure optical balancing and are detected by two identical photodiodes (Discovery Semiconductors DSC-R405ER 20 GHz Linear Balanced Photoreceivers). It is noteworthy that we carefully avoid the detector saturation at the large optical power. The difference between the photocurrent fluctuations is amplified by a homemade low-noise electronic amplifier and subsequently assessed with an electrical spectrum analyzer (ESA) (Rohde $\&$ Schwarz, 43 GHz) to obtain the radio frequency (RF) noise power spectrum. Through this balanced detection, excessive intensity noise can be substantially suppressed when compared with single photodiode detection. The common-mode rejection ratio (CMRR) consistently exceeds 30 dB across the entire frequency spectrum up to 18 GHz. 

Accurate calibration of the shot noise level, also known as the standard quantum limit (SQL), holds paramount importance in this experiment since it acts as a normalization parameter for quantifying the squeezing. In this work, we deviate from the approach in Ref.~\cite{machida1988ultrabroadband}, where the verification of the SQL was conducted using either a filtered incandescent lamp or a light-emitting diode operating at the same wavelength. Instead, we adhere to the original setup while replacing the low-noise current source with the normal pump source (Keithley 2400) \cite{mork2020squeezing}. By keeping the LO branch intact and manipulating the LO phase, we determine the SQL through this methodology. To enhance credibility, we introduce another technique by calibrating the vacuum noise level, which is performed by obstructing the LO branch. In both measurements, the RF frequency spectra of the subtracted photocurrent remain consistent, irrespective of alterations in the pump current. This coherence unequivocally signifies a high degree of correspondence between these two calibration methods, hence providing an authentic representation of the SQL.

Figure.~\ref{Fig:2} (a) showcases the measured RF noise power spectral density for the QD laser biased at 40 mA and gives a comparison between the amplitude squeezing (green curve) and anti-squeezing (black curve) with the calibrated vacuum noise level (red dashed curve). The measured variance of photocurrent fluctuations is generally a weighted combination of variances in the two field quadratures. By adjusting the relative phase difference between the signal and LO, i.e., scanning of the optical delay time, the projection angle of the two quadrature components can be selectively determined. Consequently, the squeezing spectrum is probed when the projection angle is aligned with the maximum squeezing direction, while the minimum corresponds to anti-squeezing. It is clear that the RF spectrum within the frequency range of 3 GHz to 12 GHz exhibits a significant reduction below the SQL. The most substantial noise reduction occurs near 8 GHz, showing a power reduction of 0.9$\pm 0.1$ dB (6.7\%) below the SQL. The measurement frequency range discussed here is significantly higher than the cutoff frequency of Flicker (1/f) noise. Meanwhile, at such a high pump rate, the relaxation oscillation resonance is effectively suppressed. In Figure.~\ref{Fig:2} (b), the photocurrent fluctuations obtained from the balanced receiver are presented as a function of the LO phase at 8 GHz. The measurements of the amplitude squeezing and the calibrated SQL were conducted separately during the experiment, encompassing 16 different LO phase values within a 2$\pi$ period. Nearly 2/3 of the LO phase region exhibited distinct degrees of squeezing phenomenon and the two traces in Figure.~\ref{Fig:2} (a) correspond to the cases where the LO phase equals $0.4\pi$ (anti-squeezing) and $1.2\pi$ (squeezing).

A fundamental aspect of a QD laser system with a noise-suppressed current source is its inherent capability to directly convert sub-Poissonian statistics of electrons into nonclassical photon statistics. To fully comprehend the underlying mechanisms involving carrier scattering and light-matter interaction, we employ a set of coupled stochastic differential equations that are adeptly tailored to the three-energy-level QD model \cite{wang2014phase}:
\begin{align}
    \frac{\mathrm{d} N^{RS}}{\mathrm{d} t}&=R_{\textrm{pump}}+R_{ES \to RS}-R_{RS \to ES}-R^{RS}_{\textrm{decay}}\label{Equ:1}\\
    \frac{\mathrm{d} N^{ES}}{\mathrm{d} t}&=R_{RS \to ES}+R_{GS \to ES}-R_{ES \to RS} \nonumber\\
    &-R_{ES \to GS}-R^{ES}_\textrm{decay}\label{Equ:2}\\
    \frac{\mathrm{d} N^{GS}}{\mathrm{d} t}&=R_{ES \to GS}-R_{GS \to ES}-R_\textrm{stim}-R^{GS}_\textrm{decay}\label{Equ:3}\\
    \frac{\mathrm{d} S^{GS}}{\mathrm{d} t}&=R_{stim}-R_\textrm{internal}-R_\textrm{mirror}+R_{\textrm{spon}}\label{Equ:4} 
\end{align}
where $N^i$ (i = RS, ES, GS) represents the carrier density in the reservoir states (RS), excited states (ES), and ground states (GS), respectively. $S^{GS}$ denotes the photon density in the GS, $R^{i}_\textrm{decay}$ (i = RS, ES, GS) represents the spontaneous emission rate that reduces the carrier density in each energy state. $R_{ES \to RS}$ and $R_{RS \to ES}$ describe the carrier scattering rates between RS and ES, while $R_{ES \to GS}$ and $R_{GS \to ES}$ depict the carrier scattering rates between ES and GS. $R_\textrm{stim}$ accounts for the stimulated emission rate solely on GS, and $R_\textrm{spon}$ represents the fraction of the spontaneous emission coupled into the lasing mode. Furthermore, the sum of $R_\textrm{internal}$ and $R_\textrm{mirror}$ comprises the photon decay rate, with $R_\textrm{mirror}$ the outcoupling rate via facet mirrors into the output channel and $R_\textrm{internal}$ containing other internal photon losses. The definitions of the quantities appearing on the right-hand side of Eqs.~(\ref{Equ:1})--(\ref{Equ:4}) and their corresponding parameters are provided in the Supplementary Material. The stimulated RF spectrum is depicted in Figure.~\ref{Fig:2} (c). In this approach, we derive the SQL and the amplitude squeezing independently for normal pumping and quiet pumping conditions by applying Fourier transforms of the time traces. Although a significant squeezing effect can be theoretically anticipated at low frequencies, it is often overshadowed by technical noise in practical settings \cite{machida1988ultrabroadband}. Nonetheless, in both the experiments and the simulations, a discernible signature of the amplitude squeezing is observed at relatively high frequencies. In addition, we computed the normalized second-order correlation at zero delay $g^{(2)}(0)$, as depicted in Figure.~\ref{Fig:2} (d), using the simulation data. Following the relationship $\left<\Delta S^2\right>=\left[g^{(2)}(0)-1\right]\left<S\right>^2+\left<S\right>$ \cite{mork2020squeezing}, with $S$ representing the external photon number, we observed that stimulated $g^{(2)}(0)$ rapidly converges to approximately 1 shortly after passing the laser threshold $I_{th}$. However, under normal pump conditions, $g^{(2)}(0)$ slightly exceeds 1 at $4\times I_{th}$, whereas with the quiet pump, $g^{(2)}(0)$ dips just below unity at $4\times I_{th}$ due to the relatively high average photon number. Again, we conducted corresponding $g^{(2)}(0)$ measurements at 40 mA, and the experimental results are indicated by two symbols within the same figure, supporting our simulation findings. A recent theoretical study has affirmed that a single-mode Gaussian state can exhibit a $g^{(2)}(0)<1$ \cite{olivares2018homodyning}. The spectral distribution of the field is observed to be centred around 1308.5 nm in the experiment (see Supplementary Material). Further investigations can be carried out through the reconstruction of the Wigner function \cite{smithey1993measurement}.
\begin{figure}[htbp]
\centerline{\includegraphics[width=0.5\textwidth]{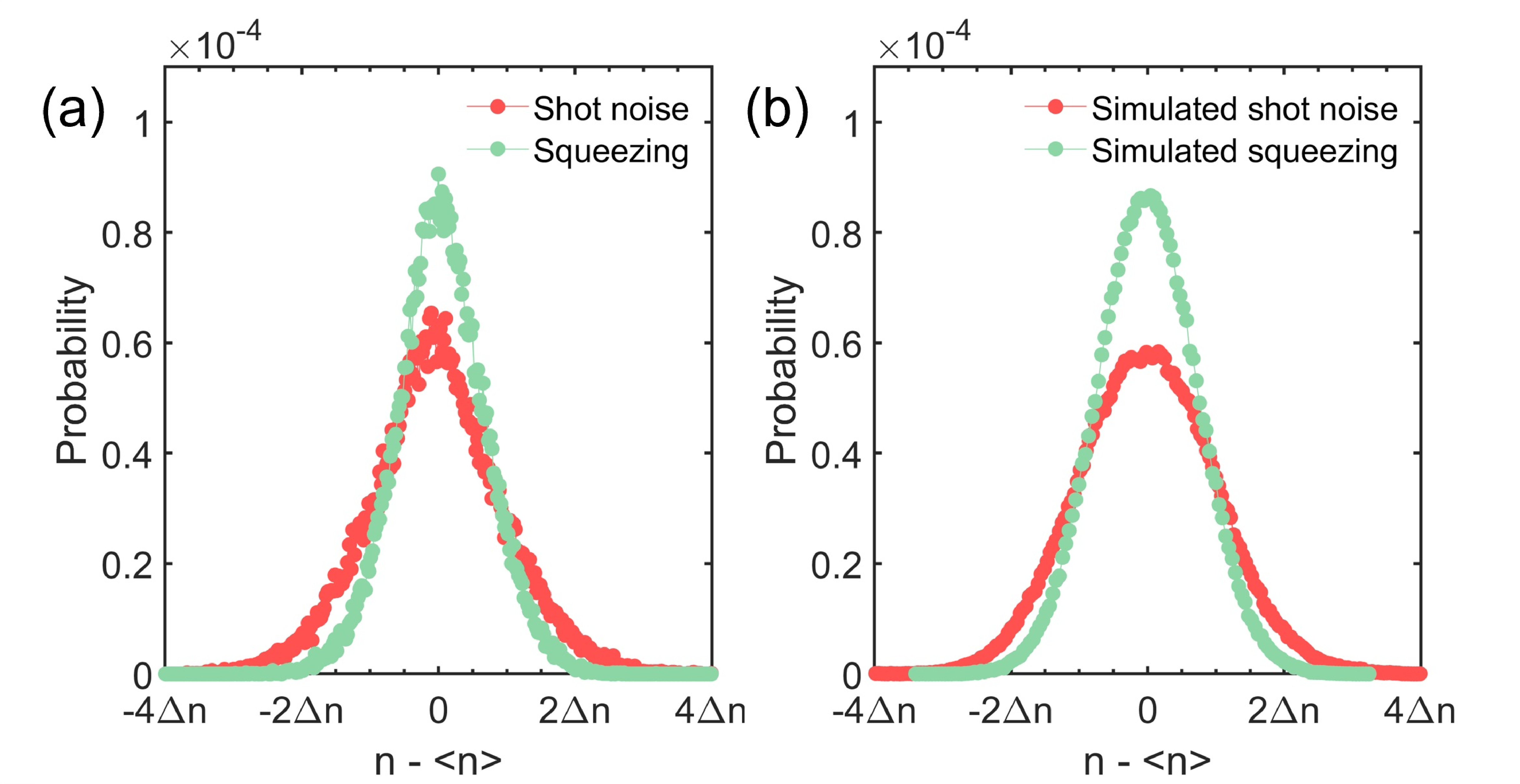}}
\caption{(a) Histogram calculated from the raw data acquired by the oscilloscope. Each time trace is averaged over 5 frames and the sampling rate is 20 Gbps. (b) Histogram computed from the simulation data, where each time trace has 2$\times 10^6$ data points, spanning a time period of 1 $\mu$s.
\label{Fig:3}}
\end{figure}

Figure.~\ref{Fig:3} provides a comparison between measured and simulated photon statistics. The photon number distributions are extracted from experimental data recorded by the oscilloscope and from stochastic simulations. In both scenarios, the histogram for shot noise confirms a Poissonian distribution. However, under quiet pumping conditions, the distribution becomes narrower, exhibiting sub-Poissonian characteristics that are confirmed by the measurement of the $g^{2}(0)$ below 1. 

Yamamoto \textit{et al.} envisaged that a DFB semiconductor laser with a high internal quantum efficiency could manifest a broad squeezing frequency bandwidth, typically exceeding 100 GHz, ultimately determined by the semiconductor laser photon lifetime ($\tau_p\sim$1 ps) \cite{yamamoto1986amplitude}. However, merely relying on a constant-current source with reduced shot noise does not unambiguously dictate an anti-correlation between successive injection events into the active layer. This is because each carrier injection constitutes a (Poisson) random point process, driven solely by the junction voltage and the junction's temperature \cite{imamog1993noise}. Nevertheless, it should be underlined that the injection rate can be influenced by the Coulomb blockade effect. For instance, the injection of a carrier results in a reduction of the junction voltage by $e/C_{\textrm{dep}}$, where $C_{\textrm{dep}}$ represents the depletion layer capacitance. This voltage reduction accordingly leads to a decrease in the carrier injection rate, establishing a negative feedback mechanism that suppresses noise in the carrier injection process. This mechanism operates successfully in the single-photon turnstile device within the mesoscopic domain \cite{imamog1994turnstile} at the low temperature. But in the macroscopic high-temperature regime, $N=(\frac{k_BT}{e})/(\frac{e}{C_{\textrm{dep}}})=k_BTC_{\textrm{dep}}/e^2$ electrons lead to a substantial junction voltage reduction equivalent to the thermal voltage $k_BT/e$, resulting in the pronounced regulation of the carrier injection rate. Considering that electrons are provided by the current source at a rate of $I/e$, the time required for $N$ electrons to be delivered is $\tau_{te}=k_BTC_{\textrm{dep}}/eI$. Therefore, this collective behaviour of numerous electrons introduces an additional limitation to the squeezing bandwidth $B$ \cite{imamog1993noise}. 
\begin{align}
    B=\frac{1}{2\pi(\tau_{te}+\tau_p)}=\frac{1}{2\pi(\frac{k_BTC_{\textrm{dep}}}{eI}+\tau_p)}
    \label{Equ:5}
\end{align}
The squeezing bandwidth is expected to be directly proportional to the current $I$ and the capacitance $C_{\textrm{dep}}$, but inversely proportional to the temperature $T$. With the specific numerical parameters outlined in Ref.~\cite{richardson1991squeezed}, where the depletion layer capacitance of the QW laser is around 280 pF, the calculated squeezing bandwidth is estimated to be around 1 GHz at 66 K. Meanwhile, QD lasers generally feature lower values of $C_{\textrm{dep}}$, sometimes as low as 3.5 pF \cite{inoue2018directly}. This would give rise to a scenario where $\tau_{te}\ll\tau_p$, resulting in a calculated squeezing bandwidth of several tens of gigahertz at room temperature, given that $\tau_p$ is in the order of a few picoseconds.

In conclusion, our work has achieved a significant milestone by successfully generating broadband amplitude-squeezed states operating at room temperature using constant-current-driven QD DFB lasers. The RF spectrum in homodyne detection has been reduced by up to 0.9 dB across a wide frequency range of 3 GHz to 12 GHz compared to the shot noise level. While the observed degree of squeezing remains modest due to certain deficiencies or photon loss within the setup, it must be pointed out that this limitation is only determined by the internal quantum efficiency of the laser itself \cite{machida1988ultrabroadband}. As a result, the prospects for further enhancing amplitude squeezing performance appear to be quite promising, considering the attainable internal quantum efficiency exceeding 90\%. This optimistic projection could be fueled by the advancements in QD laser cavity design \cite{ostrowski2023no}, the external feedback method \cite{kitching1995room} and the ongoing integration of the entire system onto a single chip \cite{tasker2021silicon}. The achievement of 85\% (-8.3 dB) amplitude squeezing through the face-to-face positioning of the detector and the laser, as demonstrated in previous research \cite{richardson1991squeezed}, provides compelling evidence for the feasibility of these improvements. In practical scenarios, a squeezing degree of 3-6 dB has proven sufficient to surpass coherent states in quantum communication \cite{waks2002security,derkach2020squeezing}. This level of squeezing is particularly productive in mitigating channel noise caused by atmospheric turbulence in free-space transmission, which could otherwise compromise the secret key rate and communication security. The findings of this study carry direct implications for diverse quantum information applications, especially in the CV QKD over free-space channels. Bright squeezed semiconductor lasers, as proposed in Ref.~\cite{hosseinidehaj2022simple}, offer perfect light sources in this context. For further insights into a simplified CV QKD configuration, refer to the Supplementary Material. Lastly, we firmly believe that this technology represents an enormous stride toward the full on-chip implementation of CV QKD protocols in the near future.

\begin{acknowledgments}
Shiyuan Zhao and Shihao Ding contributed equally to this work and jointly conceptualized the work; Shiyuan Zhao conducted the simulation, and Shihao Ding conducted the experiments. The technical contributions of Prof. Wolfgang Els{\"a}{\ss}er are gratefully acknowledged. The authors acknowledge the financial support of the Institut Mines-T{\'e}l{\'e}com and the Air Force Office of Scientific Research (AFSOR) under grant FA8655-23-1-7050. Shihao Ding’s work is also supported by the China Scholarship Council (CSC).
\end{acknowledgments}

\nocite{*}

\bibliography{apssamp}

\end{document}